\documentclass[twocolumn,superscriptaddress,prl,aps]{revtex4}
\usepackage{amsfonts,amssymb,amsmath,amsthm,graphicx,epsfig}

\begin{document}
\title{Casimir interaction of dielectric gratings.}
\author{Astrid Lambrecht}
\email{lambrecht@spectro.jussieu.fr} \affiliation{Laboratoire
Kastler Brossel, CNRS, ENS, UPMC, Campus Jussieu case 74, 75252
Paris, France }
\author{Valery N. Marachevsky}
\email{maraval@mail.ru} \affiliation{Laboratoire Kastler Brossel,
CNRS, ENS, UPMC, Campus Jussieu case 74, 75252 Paris, France }
\affiliation{V. A. Fock Institute of Physics, St. Petersburg State
University, 198504 St. Petersburg, Russia }

\begin{abstract}
We derive an exact solution for the Casimir force between two
arbitrary periodic dielectric gratings and illustrate our method by
applying it to two nanostructured silicon gratings. We also
reproduce the Casimir force gradient measured recently \cite{Chan}
between a silicon grating and a gold sphere taking into account the
material dependence of the force. We find good agreement between our
theoretical results and the measured values both in absolute force
values and the ratios between the exact force and PFA predictions.
\end{abstract}

%PACS numbers: 03.70.+k, 12.20.Ds, 42.50.Lc

\maketitle

\section{Introduction}

The availability of experimental set-ups that allow accurate
measurements of surface forces between macroscopic objects at
submicron separations has recently stimulated a renewed interest in
the Casimir effect. In 1948 H. B. G. Casimir showed that two
electrically neutral, perfectly conducting plates, placed parallel
in vacuum, modify the vacuum energy density with respect to the
unperturbed vacuum\cite{Casimir}. The vacuum energy density varies
with the separation between the mirrors and leads to the Casimir force,
which scales with the inverse of the
forth power of the mirrors separation $L$.

The Casimir force is highly versatile and tailoring it
could potentially be useful in the design and control of micro- and
nanomachines. While the material dependence of the Casimir force has
been thoroughly studied between two plane mirrors (see e.g.
\cite{Lifshitz56,Lambrecht,Mostepanenko,Pirozhenko2}), for most
other geometries exact calculations exist only for perfectly
reflecting boundaries (see e.g.\cite{Marachevsky}). If material properties are taken into
account, the shape dependence of the Casimir force is usually treated using the proximity force approximation (PFA) which amounts
to summing up contributions at different distances as if they were
independent. 
%For example the role of surface plasmons to tailor the Casimir force
%\cite{Barton,Intravaia,Henkel,Bordag} and the possible interest of
%metamaterials \cite{Henkel2,Leonhardt,Pirozhenko,Dalvit} in order to
%produce repulsive forces have been studied during the last years.

In a recent paper\cite{Chan}, Chan et al. present the first
measurement of the Casimir force between a silicon grating of high
aspect ratio and a gold sphere and demonstrate the violation of PFA in this geometry. Corresponding
calculations taking into account the periodic structure beyond PFA,
but only for perfect mirrors\cite{Emig}, turn out to lead to a too
large deviation from PFA\cite{Chan}.

In this Letter  we present the first exact calculation of the
Casimir force between gratings of arbitrary periodic structure,
where we take explicitly into account the (arbitrary) dielectric
permittivity of the material. We first present formulations for the
Casimir energy between two periodic dielectric gratings and outline
the derivation of these formulae. We then apply our formulation to
the situation of two rectangular silicon gratings and show that our
calculation yields deviations of the real force from the PFA
prediction up to $24$ percents. We also performed calculations
corresponding to the measurement by Chan et al. allowing therefore a
first quantitative theory-experiment comparison. The result taking
into account the finite conductivity gives a smaller deviation of
the exact force from the PFA prediction than the calculation for
perfect mirrors.

\section{General procedure}
\begin{figure}
\centering \includegraphics[width=8cm]{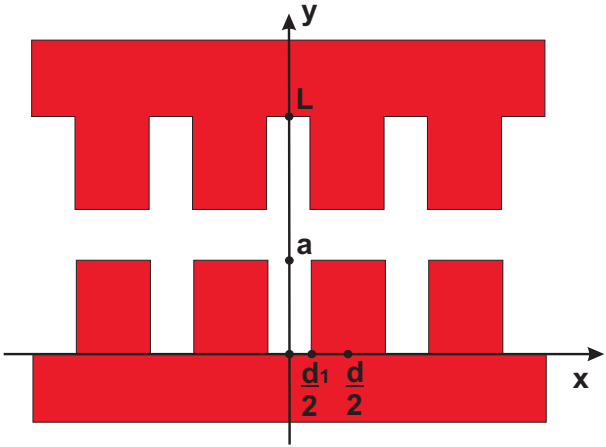}
\caption{Rectangular gratings geometry.} \label{gratings}
\end{figure}

We consider two periodic dielectric gratings of arbitrary form
separated by a vacuum slit. The special case of lamellar (or
rectangular) gratings is depicted in Fig.\ref{gratings}. The
geometrical parameters are the corrugation depth $a$, the period $d$
and the gap $d_1$. The gaps of both gratings are separated by a
distance $L$. For simplicity we will suppose the space between the
two gratings to be filled with vacuum with $\epsilon=\mu=1$.

The physical
problem is time and $z$ invariant, so electric and magnetic fields
can be written in the form:
\begin{eqnarray}
E_i (x,y,z,t) &=& E_i (x,y) \exp(ik_z z - i \omega t)  ,  \\
H_i (x,y,z,t) &=& H_i (x,y) \exp(ik_z z - i \omega t) .
\end{eqnarray}
Let us first suppose the upper grating to be absent and
consider a generalized conical diffraction problem on the lower
grating. The longitudinal components of the electromagnetic field
outside the corrugated region ($y>a$) may be written by
making use of a generalization of the Rayleigh expansion for an
incident monochromatic wave:
 \begin{eqnarray}
E_z(x,y) &=& I_p^{(e)} \exp (i \alpha_p x - i \beta_p^{(1)}y)  +
\nonumber \\
 &&\sum_{n=-\infty}^{+\infty}
R_{np}^{(e)} \exp( i \alpha_n x  + i \beta_n^{(1)} y)  , \label{Ezp} \\
H_z(x,y) &=& I_p^{(h)} \exp (i \alpha_p x - i \beta_p^{(1)}y)  +
\nonumber \\
 &&\sum_{n=-\infty}^{+\infty} R_{np}^{(h)} \exp( i
\alpha_n x  + i \beta_n^{(1)} y) ,  \label{Hzp}\\
\alpha_p &=& k_x + 2\pi p/ d  ,\quad \beta_p^{(1)2} = \omega^2 -
k_z^2 - \alpha_p^2,
\\ \alpha_n &=& k_x + 2\pi n/ d  ,\quad \beta_n^{(1)2} = \omega^2 -
k_z^2 - \alpha_n^2
\end{eqnarray}
with an integer $p$. The sums are performed over all integers $n$.
All other field components can be expressed via the longitudinal
components $E_z, H_z$.  This solution is valid outside any periodic
one-dimensional structure.

We now have to determine the coefficients $R_{np}^{(e)},
R_{np}^{(h)}$ for a specific periodic geometry profile. For this
purpose we rewrite the Maxwell equations inside the
corrugation region $0<y<a$ in the form of first order differential
equations, $\frac{\partial A}{\partial y} = M A$, where $M$ is a
square matrix of dimension $8N+4$, $A^T =(E_z, E_x, H_z, H_x)$ and
$2N+1$ is the number of Rayleigh coefficients considered in every
Rayleigh expansion. For a rectangular dielectric grating the matrix
$M$ is a constant matrix. At $y=0$ the solution has to satisfy the
following expansions, valid for $y\le 0$:
 \begin{eqnarray}
E_z(x,y) &=&  \sum_{n=-\infty}^{+\infty}
T_{np}^{(e)} \exp( i \alpha_n x  - i \beta_n^{(2)} y)  , \label{Ezt} \\
H_z(x,y) &=&  \sum_{n=-\infty}^{+\infty}
T_{np}^{(h)} \exp( i \alpha_n x  - i \beta_n^{(2)} y)   \label{Hzt} ,\\
\beta_n^{(2) 2} &=& \epsilon \mu \omega^2 - k_z^2 - \alpha_n^2.
 \end{eqnarray}
We then determine the unknown Rayleigh
coefficients  by matching the solution of equations $\frac{\partial
A}{\partial y} = M A$ inside the corrugation region with Rayleigh
expansions (\ref{Ezp}),(\ref{Hzp}) at $y=a$ and expansions
(\ref{Ezt}),(\ref{Hzt}) at $y=0$. Everywhere in the calculations we
assumed $\mu=1$.

The fields $E_z$ and $H_z$ are not decoupled for $k_z \ne
0$. This is why the reflection matrix $R_1$ for a reflection from a
lower grating can be defined as follows:
\begin{widetext}
\begin{equation}
R_1(\omega) = \begin{pmatrix} R_{n_1 q_1}^{(e)}(I_p^{(e)}=\delta_{p
q_1}, I_p^{(h)}=0  ) \qquad
&R_{n_2 q_2}^{(e)}(I_p^{(e)}=0, I_p^{(h)}=\delta_{p q_2} )   \\
R_{n_3 q_3}^{(h)}(I_p^{(e)}=\delta_{p q_3}, I_p^{(h)}=0  ) \qquad
&R_{n_4 q_4}^{(h)}(I_p^{(e)}=0, I_p^{(h)}=\delta_{p q_4} )
\end{pmatrix} . \label{R1}
\end{equation}
\end{widetext}
Performing a  change of variables $y= - y^{\prime} +L$,
$x=x^{\prime} - s  \quad (s < d)$ in $(\ref{Ezp}), (\ref{Hzp})$, it
is possible to obtain the reflection matrix $R_{2up}$ for the
reflection of an upward wave from a grating with the same profile
turned upside-down, displaced from the lower grating by $\Delta
x=s,\, \Delta y=L$. Note that for the upper grating in
Fig.\ref{gratings} the special case $s=0$ is depicted.

Up to now we considered a diffraction problem on a single grating.
In \cite{Reynaud} the Casimir energy between two bodies, the
diffraction properties of which can be described by a scattering
matrix, has been derived in plane geometries on the basis of
canonical quantization. Roughness corrections were derived on the
basis of a scattering approach in \cite{Paulo}. The path
integral method was used to obtain multipole expansion of the
Casimir energy between the two compact objects \cite{Jaffe}, exact
results in spherical geometries \cite{Jaffe, MaiaNeto} were also
derived.

We outline a novel derivation here, which can be applied to various Casimir systems. To obtain the
Casimir energy we need to determine the eigenfrequencies of all
stationary solutions of the generalized diffraction problem of
subsequent diffraction of the electromagnetic field on two periodic
gratings separated by a gap-gap distance $L$. These eigenfrequencies
can be summed up by making use of an argument principle, which
states:
\begin{equation}
\frac{1}{2\pi i} \oint \phi(\omega) \frac{d}{d\omega} \ln f(\omega)
d\omega = \sum \phi (\omega_0) -\sum \phi(\omega_\infty) ,
\label{arg}
\end{equation}
where $\omega_0$ are zeroes and $\omega_\infty$ are poles of the
function $f(\omega)$ inside the contour of integration. Degenerate
eigenvalues are summed over according to their multiplicities. For
the Casimir energy we have $\phi(\omega) = \hbar\omega/2$. The
equation for eigenfrequencies of the corresponding problem of
classical electrodynamics is $f(\omega)=0$.

Consider first the plane-plane geometry when two dielectric parallel slabs
(slab $1$: $y<0$, slab $2$: $y>L$) are separated by a vacuum slit
($0<y<L$). In this case $TE$ and $TM$
modes do not couple. The equation for $TE$ eigenfrequencies is:
\begin{equation}
f(\omega)= 1-r_{1 TE} (\omega) r_{2 TE up} (\omega) = 0 .
\label{char}
\end{equation}
Here $r_{1 TE}(\omega)$ is  the reflection coefficient  of a downward
plane wave which reflects on a dielectric surface of slab $1$ at
$y=0$, while $r_{2 TE up} (\omega)$ is the reflection coefficient of an
upward plane wave which reflects on a dielectric surface of slab
$2$ at $y=L$. One can deduce from Maxwell equations that $r_{2 TE
up}(\omega)= r_{2 TE}(\omega) \exp(2ik_y L)$ ($r_{2 TE}(\omega)$ is
a reflection coefficient of a downward $TE$ plane wave which
reflects on a dielectric slab $2$ now located at $y<0$). From
(\ref{char}) and the analogous equation for $TM$ modes one immediately
obtains the Lifshitz formula by making use of the argument principle
(\ref{arg}).

For two periodic dielectrics separated by a vacuum slit one has to
consider a reflection of downward and upward waves from a unit cell
$0<k_x<2\pi/d$. Due to the structure of the surface, $TE$ and $TM$
modes do not decouple anymore, but they are coupled by the
diffraction process. The equation for normal modes states:
\begin{equation}
R_{1} (\omega_i) R_{2up}(\omega_i) \psi_i = \psi_i, \label{bound}
\end{equation}
where $\psi_i$ is an eigenvector describing the normal mode with a
frequency $\omega_i$. Instead of equation (\ref{char}) one obtains
from (\ref{bound}) the following condition for eigenfrequencies:
\begin{equation}
\det(I - R_{1} (\omega) R_{2up}(\omega)) = 0 .\label{EIG}
\end{equation}
For every $k_x, k_z$ the solution of (\ref{EIG}) yields  possible
eigenfrequencies $\omega_i$ of the solutions of Maxwell equations
that should be substituted into the definition of the Casimir energy
$E = \sum_i \hbar \omega_i/2$. These solutions should tend to zero
for $y \to \pm\infty$. The summation over the eigenfrequencies is
performed by making use of the argument principle (\ref{arg}), which
yields the Casimir energy of two parallel gratings on a "unit cell"
of period $d$ and unit length in $z$ direction:
\begin{align}
E = &\frac{\hbar c \: d}{(2 \pi)^3} \int_0^{+\infty} d \omega
\int_{-\infty}^{+\infty} d k_z  \int_0^{\frac{2\pi}{d}} d k_x \quad
 \nonumber \\ &\ln {\rm det} \Bigl(I - R_1 (i\omega) R_{2up}(i \omega) \Bigr) ,
\label{EC}
\end{align}
$c$ is the speed of light in vacuum. This is an exact expression
valid for two arbitrary periodic dielectric gratings separated by a
vacuum slit. It can be applied to calculate the Casimir energy of
any parallel periodic gratings made of a material described by a
dielectric function, with surface corrugations of arbitrary
geometry.

Consider the particular case $s=0$, depicted in
Fig.\ref{gratings}. From the derivation sketched above it follows
that
\begin{equation}
R_{2up}(i\omega)=K(i\omega) R_{2}(i\omega) K(i\omega) , \label{RR}
\end{equation}
where $K(i\omega)$ is a diagonal $2(2N+1)$ matrix of the form:
\begin{equation}
K(i\omega) = \begin{pmatrix} G &  0  \\
0  &  G
\end{pmatrix},  \label{K}
\end{equation}
with matrix elements $e^{-L\sqrt{\omega^2+k_z^2+(k_x+\frac{2\pi
m}{d})^2}} \quad (m=-N\ldots N)$ on a main diagonal of a matrix $G$.
Note that in all Rayleigh expansions the Fourier basis is taken
symmetrically around $m=0$. When changing the maximum value of
$m$ from $N-1$ to $N$, each Rayleigh coefficient $R_{Np} (i\omega)$
appearing in the reflection matrices is multiplied by a factor 
$\simeq e^{- 2\pi N L/d}$ coming from the matrix $K(i\omega)$. As a consequence, when 
$2 \pi N L/d \gg 1$ is satisfied, the contribution of the
coefficients $R_{Np}(i\omega)$ is suppressed exponentially.
Therefore for large enough $N$ changing $N$ has only a little impact
on the final result.

\section{Rectangular gratings}

\begin{figure}
\centering \includegraphics[width=9cm]{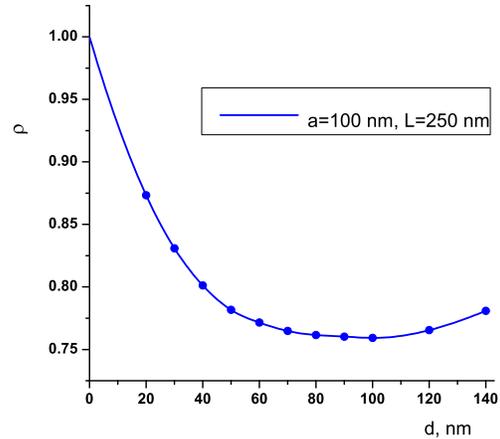}
\caption{Casimir force normalized by its PFA value for two Si
gratings with $a=100$nm and $d_1=\frac{d}{2}$ as a function of $d$
at a fixed distance $L=250$nm.} \label{FoverFPFA-A100L250_d}
\end{figure}

We have numerically calculated the exact Casimir force for two
rectangular gratings at zero temperature in the geometry of
Fig.\ref{gratings} for silicon for different values of $d$, $d1=d/2$
and $a=100$ nm by making use of the formulas (\ref{EC}, \ref{RR},
\ref{K}) and a Drude-Lorentz model for the dielectric permittivity of
intrinsic silicon \cite{Pirozhenko2}. We compare our
exact results of the Casimir force for different values of $d$ to
the PFA results. Calculated with the proximity force approximation,
the Casimir force between the two gratings is just the geometric sum
of two contributions corresponding to the Casimir force between two
plates $F_{PP}$ at distances $L$ and $L-2a$, that is
$F_\textrm{PFA}= \frac{1}{2}(F_{PP}(L) + F_{PP}(L-2a))$. In
particular it is independent of the corrugation period $d$. To
assess quantitatively the validity of PFA, we plot the dimensionless
quantity $\rho=\frac{F}{F_\textrm{PFA}}$ \cite{MaiaNeto}. The ratio
is presented on Fig.\ref{FoverFPFA-A100L250_d}. 
Exact and PFA
results differ for silicon by up to $24$ percents for a corrugation
period of 100nm and the PFA violation could thus be demonstrated
experimentally. We recover the PFA result in two limiting cases, for
a vanishing corrugation period and for very large corrugation
periods. In between the exact result for the Casimir force is always
\textit{smaller} than the PFA prediction, in contrast to
calculations for perfect conductors, where the resulting force is
always larger than the PFA prediction.

%\begin{equation}
%\sum_i \frac{\omega_i}{2}=\frac{1}{2\pi} \int_{-\infty}^{+\infty}
%\ln D(i\omega),  \quad \text{where} \:\: D(\omega_i)=0
%\end{equation}

\begin{figure}
\centering \includegraphics[width=9cm]{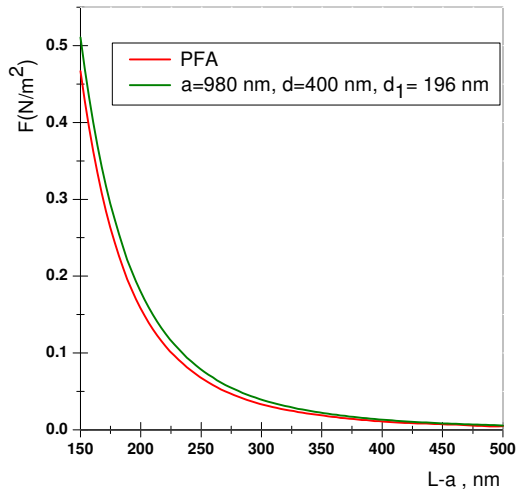} \caption{Casimir
force between a Si grating and a gold plate as a function of
distance.} \label{FChan}
\end{figure}

We will now apply our method to the recent experiment by Chan et
al.\cite{Chan}, who measured the Casimir force gradient between a
silicon grating with nanostructured trenches and a gold sphere of
radius $R=50\mu$m. The force gradient $F^{\prime}_{PS}$ between a
sphere of radius $R$ and a plate can be expressed via the force
$F_{PP}$ in the plane-plane configuration as $F^{\prime}_{PS}= 2\pi
R F_{PP}$. This is why we show in figure \ref{FChan} the zero
temperature result for the absolute force values evaluated for a
grating with the experimental parameters $a=980$nm, $d=400$nm,
$d_1=196$nm placed in front of a gold plate (we used a plasma model
with a plasma frequency $\omega_p=9 eV$ for gold
and a Drude-Lorentz model for intrinsic silicon \cite{Pirozhenko2}).

\begin{figure}
\centering \includegraphics[width=9cm]{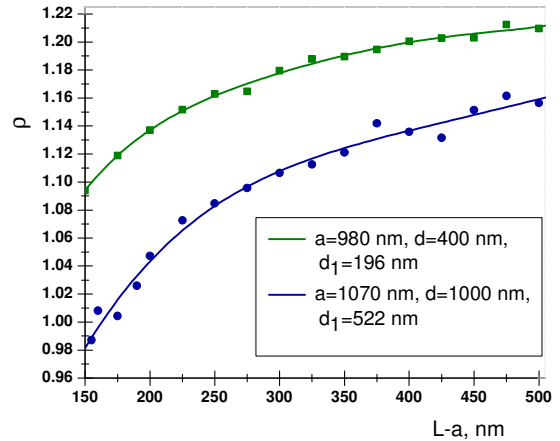}
\caption{Casimir force normalized by the PFA value between a Si
grating and a gold plate as a function of distance for two different
gratings. Solid curves are calculated by making use of the least
square method from the theoretical points on the figure.}
\label{rhoChan}
\end{figure}
From our calculation we obtain a force $F_{PP}=0.51$N/m$^2$ for a
plate separation of 150nm. With the experimental parameters this
leads to a prediction for the Casimir force gradient of
$F^{\prime}=160.8, 56.4, 24.6 $ pN/$\mu$m at respectively $L-a=150,
200, 250$ nm. The absolute values of the force are thus in good
agreement with the measured values depicted in Fig.3c of
\cite{Chan}.

We finally present ratios of our results for the force to the
predictions of PFA for two different gratings. Figure \ref{rhoChan}
shows $\rho$ as a function of $L-a$ for two gratings corresponding
to the experiment with $a=980$nm, $d=400$nm, $d_1=196$nm (green
line) and $a=1070$nm, $d=1000$nm, $d_1=522$nm (blue line) and gives reasonable agreement with experimental points and the fit 
in Fig.~3d of \cite{Chan}. 

The fact that the perfect conductor model fails might be due to the
influence of surface plasmons, as the grating affects their
dispersion relation. Surface plasmons contribute essentially and at
all distances to the Casimir force \cite{Barton,Intravaia,Henkel,
Bordag}, the Casimir force thus has to change considerably when
structured surfaces are considered. These changes are not visible in
a perfect conductor model which ignores the existence of surface
plasmons.

\begin{acknowledgments}
We would like to thank Serge Reynaud for helpful discussions.
This work was supported by the French National Research Agency (ANR)
through grant n° ANR-06-NANO-062 - MONACO project. V.M. also
acknowledges financial support from grants RNP $2.1.1.1112$, RFBR
$07-01-00692-a$, SS$.5538.2006.2$.
\end{acknowledgments}

\end{document}